\documentclass[12pt]{article}

\usepackage{amsmath,amssymb, amsthm}

\title{Born's rule from measurements of classical signals by threshold detectors which are properly calibrated}

\author{Andrei Khrennikov\\
International Center for Mathematical Modelling
\\in Physics and Cognitive Sciences\\
Linnaeus University,  V\"axj\"o, S-35195, Sweden\\
Andrei.Khrennikov@lnu.se\\
tel. 46-470708790}

\begin{document}

\maketitle

\abstract{The very old problem of the statistical content of quantum mechanics (QM) is studied in a novel framework. 
The Born's rule (one of the basic postulates of QM) is derived from theory of classical random signals. We present a measurement scheme which transforms continuous signals into discrete clicks and reproduces
the Born's rule. This is the sheme of threshold type detection. Calibration of detectors plays a crucial role. }

\section{Introduction}

In this paper we stress the role of detectors and more generally the detection procedure in the debate on completeness of 
QM. This debate was originated in the famous paper of Einstein, Podolsky, and Rosen \cite{E} in 1933
and it still attracts a lot of attention of experts in quantum foundations and quantum information (including 
researchers working on applied issues and experimentalists, especially in quantum optics), see, e.g., 
\cite{PL}--\cite{Greg2}.  Surprisingly the role of detectors and the detection procedure in  the problem of 
hidden variables has not yet been properly analysed. Typically, one points to detectors' inefficiency loophole
in experimental tests of Bell's inequality, e.g., \cite{PER}--\cite{LARS} or generally unfair sampling\cite{AD1}
but not more. And in such formal treatement it is considered as merely a technological problem which cannot be reposible
for nonclassicality of statitsical data from quantum experiments.\footnote{For example, Alain Aspect told to the author that 
he does not consider the problem of detectors' inefficiency as a problem of the fundamental value for quantum foundations
(private communication, conference ``Foundations of Probability and Physics-4'', V\"axj\"o 2007). Therefore he was not sure that the quantum community has to put resources to close this loophole.} However, 
more detailed study, e.g., \cite{AK}, of what happens in real laboratories, e.g.,  \cite{ASP}--\cite{WEIHS1}, 
demonstrated that the process of detection and real experimental routine cannot be regarded as simply 
technicalities. It seems that they play an important role in production of nonclassical statistics
 (and the author guesses -- the crucial role). 

In this paper we show that it is really the case:
 the basic postulate of QM, Born's rule, can be derived by taking into account the presence
 of detection threshold and calibration of detectors. The latter is crucial to exclude double clicks
 in ``single particle'' detection, cf. the experiment of Grangier et al.\cite{Grangier}, \cite{Grangier1}. In fact, all existing 
detectors -- photomultipliers tubes  -- PMTs,
avalanche photodiodes -- APDs, visible light counters -- VLPCs, supeconducting bolometers -- SPD 
are of such a type and all detectors have to be calibrated, see, e.g., \cite{MS}--\cite{MS1}. 
 
 The classical (``prequantum'')  model under consideration is of the wave type: classical fields are selected
 as hidden variables. (So, the state space of hidden variables, the $L_2({\bf R}^3)$-space, is infinite 
dimensional.) This is  {\it prequantum classical statisstical field theory} (PCSFT) which was recently developed by the author and collaborators \cite{KH1}--\cite{W1}. 

We state again that the field-type prequantum models are quite popular among those who question the 
completeness of QM, e.g., \cite{DBR}--\cite{Theo}. At the very beginning of the quantum theory, Schr\"odinger was sure 
that he simply discovered a new kind of wave mechanics\cite{SCHR0}. He interpreted the wave function of electron $\psi(x)$
(not normalized!) as the density of electron's charge. Hence, the quantity
\begin{equation}
\label{SCG}
q(V)= e \int_V \vert \psi(x) \vert^2 d x
\end{equation}
is the electron charge in the domain $V.$ Here $e$ is the elementary electron charge.
Shr\"odinger derived his equation starting with De Broglie's idea to associate a wave with 
any quantum particle\cite{DBR}. Later De Brogle developed his {\it Double Solution  Theory}\cite{DBR1} 
in which the wave function was interpreted as a {\it physical wave.} (Particles appear as singularities
in those waves.) His approach was essentially elaborated by Bohm -- to the pilot wave prequantum model, e.g.,
\cite{BH}.
We remark that our own approach PCSFT is closer to first attempts of Schr\"odinger to proceed in a purely 
wave framework, i.e., to consider e.g. the wave of electron charge, not electron-particle. Oppoiste to 
early Schr\"odinger and our approach, Bohmian mechanics is not aimed towards a purely wave model of
prequantum reality. PCSFT is close to the idea of Einstein to create a purely field model of physical reality,
see, e.g., Einstein and Infeld\cite{EI}. Among other wave-like prequantum models, we also mention stochastic electrodynamics  and semiclassical approach \cite{LAMBA}--\cite{Theo}. 

In this paper we show that PCSFT, a model of classical field type, can be completed by detection theory 
which produces Born's rule for probabilities of clicks of detectors. This is an important improvement of PCSFT.
In its previous version \cite{KH1}--\cite{W1}, it reproduced quantum averages and correlations, but as averages and correlations
of (continuous) {\it intensities} of classical signals. Transition from signal inetensity to probability of discrete clicks
is a real step towards quantum-like representation of the classical field theory. We discuss this issue more carefully. 

Consider the classical electromagnetic field $(E(x), B(x)),$ where $E(x)$ and $B(x)$ are electric and magnetic 
fields, respectively. To make the analogy with QM closer, we use Riemann-Silberstein representation:
$$
\psi(x)= E(x)+i B(x).
$$
(We remark that this complex function is not $L_2$-normalized.) Then the energy of the field concentrated 
in the domain $V$ is given by 
\begin{equation}
\label{SCGa}
{\cal E}(V)= \int_V \vert \psi(x) \vert^2 d x,
\end{equation}
cf. (\ref{SCG}).
Of course, the formula (\ref{SCGa}) as well as (\ref{SCG}) remind very much Born's rule
\begin{equation}
\label{SCGb}
P(x\in V)= \int_V \vert \Psi(x) \vert^2 d x,
\end{equation}
where we use the symbol $\Psi$ to distinguish the normalized wave function from non-normalized physical field.
The first formal difference between (\ref{SCGa}) and (\ref{SCGb}) is  aforementioned normalization. However,
 this is a minory problem. Consider the total energy of the calssical electromagnetic field
\begin{equation}
\label{SCGd}
{\cal E}= \Vert \psi\Vert^2=\int_{{\bf R}^3} \vert \psi(x) \vert^2 d x,
\end{equation}
Then the realative portion of electromagnetic energy concentrated in the domain $V$ is given by 
\begin{equation}
\label{SCGE}
\frac{{\cal E}(V)}{{\cal E}}= \frac{1}{ \Vert \psi\Vert^2} \int_V \vert \psi(x) \vert^2 d x=
\int_V \vert \Psi(x) \vert^2 d x.
\end{equation}
where 
$$
\Psi(x)= \frac{\psi(x)}{\Vert \psi\Vert}= \frac{E(x)+ B(x)}{\sqrt{E^2(x)+B^2(x)}},
$$
is the ``normalized electromagnetic field''. Therefore it is natural to guess that Born's rule (\ref{SCGb})
for probability of detection reflects in some way the rule for relative energy concentration.\footnote{According
to quantum folklore, Max Born invented his rule starting with the Schr\"odinger's rule (\ref{SCG}) for the
electric charge density.} The main problem is that Born's rule (\ref{SCGb}) is about {\it probability} of discrete
clicks and the rule (\ref{SCGd}) is about {\it intensity} of continuous  signal. In this paper we show that  
Born's rule (\ref{SCGb}) can be really derived from (\ref{SCGE}) and not only for the electromagnetic field.

Besides aforementioned restrictions on the detection procedure, the derivation is based on an important 
mathematical assumption on prequantum random fields. They have to be {\it ergodic.} Ergodicity 
provides a possibility to reduce ensemble averages to time averages and this is the cornerstone of
our derivation.

Some readers might be not so much interested in a new model with hidden variables. In such a case the paper can be considered as a test of boundaries of simulation of quantum statistical data with the aid  of classical signal theory.

Therefore we do not start directly with the prequantum model -- PCSFT. We start with classical theory 
of random signals,  section \ref{CLS}, cf. \cite{S1}, \cite{S2}. Then we derive Born's rule for 
threshold type detectors, section \ref{mesj} and represent the calibration procedure in this framework,
section \ref{CAL}. (We emphasize that in our model the probability of detection is not changed by calibration.)
In section  \ref{CAL1} we demonstrate that one can exclude double clicks in experiments with ``single
particle detection'' -- as in the two slit experiment, in context such that two detectors are placed
behind two slits (and single photon source). Here the proper calibration plays a crucial role. 
Finally, in section \ref{QP} we present our model with hidden variables PCSFT and then couple it to the detection
scheme elaborated in sections \ref{mesj}, \ref{CAL}, \ref{CAL1}.

We point to coupling with the operational approach to QM, cf., e.g., \cite{Ozawa}--\cite{MDA1}. In fact, we present the operational approach to measurement of classical random signals.
This approach reprodcues quantum probabilities. (Of course, we are well aware that the majority of 
the operational community stays firmly on the Copenhagen's position, i.e., that QM is complete.)

We also remark that recently the question of validity of Born's rule has attracted a lot of attention. We can mention the experimental test \cite{Sorkin},\cite{Sorkin1} of Sorkin's model (see, e.g., \cite{Sorkin} for 
brief presentation) with violation of Born's rule for triple-slit experiment and ideas on quantum supercomputing
based on quantum-like models violating Born's rule \cite{Howard}. By PCSFT (completed with measurement theory developed in this paper) validity of Born's
rule (\ref{SCGb}) in QM is a consequence of the formula (\ref{SCGa}) (in fact, its ``normalized version'' (\ref{SCGE})), i.e., the fact that 
in classical signal theory the energy at the point $x\in {\bf R}^3$ is proportional to {\it squared field} 
\begin{equation}
\label{SCGE1}
{\cal E}(x) \equiv {\cal E}_2(x)= \vert \psi(x)\vert^2.
\end{equation}
Of course, the assumption that {\it detectors integrate the energy density} plays a crucial role. If some kind of detector also were able 
to integrate nonquadratic field nonlinearities, then Born's rule can be violated (in measurements with detectors of such a type).
For example \cite{KH1p}, suppose that a detector can also integrate the contribution of the quantity
\begin{equation}
\label{SCGE2}
{\cal E}_4(x)= \vert \psi(x)\vert^4.
\end{equation}
Hence, totally the quantity
\begin{equation}
\label{SCGE2}
{\cal E}_{2,4}(x)= \vert \psi(x)\vert^2 + \alpha \vert \psi(x)\vert^4,
\end{equation}
where $\alpha$ is a small parameter, contributes to a click. Then Born's rule has to be violated:
\begin{equation}
\label{SCGE3}
 P(x\in V)= \int_V \vert \Psi(x) \vert^2 d x + \alpha F(\vert \Psi\vert^2, \vert \Psi\vert^4).
\end{equation}
In \cite{KH1p} the addtitional contribution $F(\vert  \Psi\vert^2, \vert \Psi \vert^4)$ has been calculated.
The small parameter $\alpha$ determines the scale of prequantum fluctuations, so if a detector even were able 
to integrate fourth-order nonlinearity, the corresponding perturbation would be very small.

Thus, by PCSFT-approach to QM Born's rule might be in principle violated. However, this would not have a fundamental value
for theory, since such violation would be coupled solely to the detection procedure and not properties of the 
prequantum random fields.

\section{Classical random signals: ensemble and time representations of averages}  
\label{CLS}

The state space of classical signal theory  is the $L_2$-space $$H=L_2({\bf R}^3).$$ Elements of 
$H$ are classical fields $\phi: {\bf R}^3 \to {\bf C}^n.$ We consider complex valued fields. (For
example, for the classical electromagnetic field we use Riemann-Silberstein representation,
$\phi(x) = E(x) +i B(x).$) 

A random field (signal) is a field (signal) depending on a random parameter $\omega, 
\phi(x, \omega).$ In the measure-theoretic framework (Kolmogorov, 1933) it is represented as 
$H$-valued random variable, $\omega \to \phi(\omega)\in H.$ Its probability distribution is denoted by the symbol 
$\mu$ on $H.$  

Consider functionals of fields, $$f: H \to {\bf C}, \phi \to f(\phi).$$ These are physical observables for classical signals.
For example, the energy of the classical electromagnetic field is geven by the quadratic functional
$$
f(\phi)\equiv f(E, B)= \int_{{\bf R}^3} \vert \phi(x) \vert^2 dx=  \int_{{\bf R}^3} (E^2(x) + B^2(x)) dx.
$$
The average of an  observable can be written as the integral over the space of fields
$$
\langle f\rangle = \int_H f(\phi) d \mu(\phi).
$$
To find $\langle f\rangle,$ we consider an ensemble (in theory infinite) of realizations of the 
random field and calculate the average of $f(\phi)$ with respect to this ensemble. This measure-theoretic 
(ensemble) representation is very convenient in theoretical considerations \cite{S1}, \cite{S2}. However, in practice
we never produce an ensemble of different realizations of a signal.  Instead of this, we have a single time
dependent realization of a signal, $\phi(s,x).$ It is measured at different instances of time. Finally, we calculate 
the time average. The latter is given by 
\begin{equation}
\label{eqg1}
\bar{f}= \lim_{\Delta \to \infty} \frac{1}{\Delta} \int_0^\Delta f(\phi(s)) ds.
\end{equation}
In classical signal theory \cite{S1}, \cite{S2} the ensemble and time averages are coupled by
the {\it ergodicity assumption.} Under this assumption we obtain that 
\begin{equation}
\label{eqg2}
\bar{f}= \langle f \rangle,
\end{equation}
i.e., 
\begin{equation}
\label{eqg3}
\int_H f(\phi) d \mu(\phi) = \lim_{\Delta \to \infty} \frac{1}{\Delta}\int_0^\Delta f(\phi(s)) ds \approx 
\frac{1}{\Delta}\int_0^\Delta f(\phi(s)) ds,
\end{equation}
foe sufficiently large $\Delta.$ 

In coming consideration we shall operate only with observables given by quadratic functionals of 
classical signals
\begin{equation}
\label{eqg4}
\phi \to f_A(\phi) = \langle \widehat{A} \phi, \phi \rangle,
\end{equation}
where $\widehat{A}$ is a self-adjoint operator.  Moreover, to describe a procedure of the position detection we  need only functionals of the form
\begin{equation}
\label{eqg5}
\phi \to \vert \phi(x_0)\vert^2,
\end{equation}
where $x_0 \in {\bf R}^3$ is a fixed point which determines the quadratic functional
(later $x_0$ will be considered as the position of a detector). 

In what follows we consider only random signals with covariance operators of the type 
\begin{equation}
\label{eqg41}
D_\psi= \vert \psi\rangle \langle \psi \vert,
\end{equation}
 where $\psi \in H$ is arbitrary vector 
(i.e., it need not be normalized by 1).

We remark that we still proceed in the framework of classical signal theory, i.e., without any relation to QM. We simply 
consider a special class of classical random signals characterized by the form of the covariance operator. For example, 
we can consider Gaussian beams. For a Gaussian signal (with zero mean value), the form (\ref{eqg41}) of the covariance operator 
implies that this signal is concentrated on one dimensional subspace of the $L_2$-space based on the vector $\psi.$ We also
state again that for Gaussian signals the covariance operator determines signal uniquely (for the fixed mean value).

For such probability measure $\mu\equiv \mu_\psi,$ 
\begin{equation}
\label{EQj1}
\langle f_{x_0}\rangle = \int_H \vert \phi(x_0)\vert^2 d\mu_\psi(\phi) = \vert \psi(x_0)\vert^2.
\end{equation}
And under the assumption of ergodicity, we obtain
\begin{equation}
\label{EQj2}
  \vert \psi(x_0)\vert^2= \lim_{\Delta \to \infty} \frac{1}{\Delta} \int_0^\Delta \vert \phi(s, x_0)\vert^2 ds 
\approx \frac{1}{\Delta} \int_0^\Delta \vert \phi(s, x_0)\vert^2 ds,
\end{equation}
for sufficiently large $\Delta.$
Consider the functional 
\begin{equation}
\label{EQj3}
\pi(\phi) = \Vert \phi \Vert^2= \int_{{\bf R}^3} \vert \phi(x)\vert^2  dx.
\end{equation}
In PCSFT it represents the total energy of a signal. We find its average. In general,
\begin{equation}
\label{EQj4}
\langle \pi\rangle = \int_H \pi(\phi) d\mu(\phi) = \rm{Tr} D_\mu.
\end{equation}
In particular, for $\mu=\mu_\psi,$ 
\begin{equation}
\label{EQj5}
\langle \pi\rangle = \int_H \pi(\phi) d\mu_\psi(\phi) = \Vert \psi \Vert^2.
\end{equation}
By ergodicity
\begin{equation}
\label{EQj6}
\langle \pi\rangle = \Vert \psi \Vert^2 = \lim_{\Delta \to \infty} \frac{1}{\Delta} \int_0^\Delta \Vert \phi(s)\Vert^2 ds
\approx \frac{1}{\Delta} \int_0^\Delta ds \int_{{\bf R}^3}  dx \vert \phi(s,x)\vert^2,
\end{equation}
for sufficiently large $\Delta.$

If, as usual in signal theory, the quantity $\vert \phi(s,x)\vert^2$ has the physical dimension 
of the energy density, i.e., energy/volume, then by selecting some unit of time denoted $\gamma$
we can interpret the quantity 
 \begin{equation}
\label{EQj7}
\frac{1}{\gamma} \int_0^\Delta \vert \phi(s, x_0)\vert^2 ds dV,
\end{equation}
as the energy which can be collected in the volume $dV$ during the time interval $\Delta$ (from the random 
signal $\phi(s) \in H).$ In the same way 
\begin{equation}
\label{EQj8}
\frac{1}{\gamma} \int_0^\Delta ds \int_{{\bf R}^3} dx \vert \phi(s, x)\vert^2,
\end{equation}
is the total energy which can be collected during the time interval $\Delta.$ 
Its time average can be represented in the form (\ref{EQj6}).

\section{Discrete-counts model for detection of classical random signals}
\label{mesj}

We consider the following model of detector's functioning. Its basic parameter is detection 
threshold energy $\epsilon\equiv \epsilon_{\rm{click}}.$ The detector under consideration clicks
after it has been collected the energy 
\begin{equation}
\label{EQj9}
E_{\rm{collected}} \approx \epsilon.
\end{equation}
Such a detector is calibrated to work in accordance with (\ref{EQj9}). Clicks with energies deviating from 
$\epsilon$ are discarded. Detectors are calibrated for a class of signals and $\epsilon$ is selected
as 
\begin{equation}
\label{EQj10}
\epsilon \approx \langle \pi\rangle = \Vert \psi \Vert^2,
\end{equation}
the average energy of a signal.

Select $\gamma$ as e.g. one second. Consider such a detector located in small volume $dV$ around a point  $x_0 \in {\bf R}^3.$ 
In average it clicks each  $\Delta$ seconds, where $\Delta$ is determined form the approximative equality
\begin{equation}
\label{EQj11}
\frac{1}{\gamma} \int_0^\Delta \vert \phi(s, x_0)\vert^2 ds dV\approx \epsilon,
\end{equation}
or 
\begin{equation}
\label{EQj12}
\frac{\Delta}{\gamma} \Big( \frac{1}{\Delta} \int_0^\Delta \vert \phi(s, x_0)\vert^2 ds\Big) dV\approx \Vert \psi \Vert^2,
\end{equation}
or 
\begin{equation}
\label{EQj13}
\frac{\Delta}{\gamma} \vert \psi(x_0)\vert^2 dV \approx \Vert \psi \Vert^2.
\end{equation}

Let us introduce the normalized function $$\Psi(x)= \psi(x) /\Vert \psi \Vert,$$ i.e., $\Vert \Psi \Vert^2=1.$
Then (\ref{EQj13}) can be rewritten as 
\begin{equation}
\label{EQj14}
\frac{\Delta}{\gamma} \vert \Psi(x_0)\vert^2 dV  \approx 1.
\end{equation}
Thus at the point $x_0$ such a detector clicks (in average) with the frequency
\begin{equation}
\label{EQj15}
\lambda(x_0) = \frac{\gamma}{\Delta} \approx \vert \Psi(x_0)\vert^2 dV. 
\end{equation}
This frequency of clicks coincides with the probability of detection at the point $x_0.$  

Consider a large interval of time, say $T.$ The number of clicks at $x_0$ during this interval 
is given by
\begin{equation}
\label{EQj16}
n_T(x_0) = \frac{T}{\Delta} \approx\frac{1}{\gamma}  \vert \Psi(x_0)\vert^2 dV. 
\end{equation}
The same formula is valid for any point $x \in {\bf R}^3.$ Hence, the probability of detection at $x_0$
is 
\begin{equation}
\label{EQj17}
P(x_0) = \frac{n_T(x_0)}{\int n_T(x) dx} \approx \frac{\vert \Psi(x_0)\vert ^2 dV}{\int \vert \Psi(x)\vert^2 dx}=
\vert \Psi(x_0)\vert ^2 dV.
\end{equation}
Here $\Psi(x)$ is a kind of the wave function, a normalized vector of the $L_2$-space. (We state again that we still consider
just classical signal theory.)

\medskip

{\bf Conclusion.} {\it Born's rule is valid for probabilities of ``discretized detection'' of classical
random signals under the following assumptions:

(a) ergodicity;

(b) a detector clicks after it ``has eaten'' approximately a portion of energy $\epsilon;$

(c) a detector is calibrated in such a way that this click threshold $\epsilon$ equals to the 
average energy of a signal (or more generally a class of signals);

(d) the energy is collected by this detector through time integration of signal's energy;

(e) the interval of integration $\Delta$  is long enough from the viewpoint of the 
internal time scale of a signal.}

\medskip

The assumption (e) is necessary to match (a). We remark that the 
internal time scale of a signal, i.e., the scale of its random fluctuations, 
 has to be distinguished from the time scale of macroscopic measurement (observer's 
 time scale). The former is essentially finer than the latter.
 
 We presented a natural scheme of discrete detections which is based on time 
integration of signal's energy by a detector. Calibration of the detector  plays a
crucial role. This scheme applied to classical random signals reproduces Born's rule 
for {\it discrete clicks.}

\section{Invariance of Born's rule with respect to calibration of detectors}
\label{CAL}

Let us repeat the previous considerations by scaling the detection threshold. We now put
\begin{equation}
\label{TRESH}
\epsilon = C \Vert \Psi \Vert^2,
\end{equation}
where $C>0$ is an arbitrary scaling constant. We interpret scaling (\ref{TRESH}) as {\it calibration}
of detector.

We now repeat considerations of section \ref{mesj} with this scaled threshold. For
example, the analog of the equality (\ref{EQj13})  has the form
\begin{equation}
\label{EQj13C}
\frac{\Delta_C}{\gamma} \vert \psi(x_0)\vert^2 dV \approx C \Vert \psi \Vert^2,
\end{equation}
or for the normalized function $\Psi(x)= \psi(x) /\Vert \psi \Vert,$ 
\begin{equation}
\label{EQj14C}
\frac{\Delta_C}{\gamma} \vert \Psi(x_0)\vert^2 dV  \approx C.
\end{equation}
Thus at the point $x_0$ such a detector clicks (in average) with the frequency
\begin{equation}
\label{EQj15C}
\lambda_C(x_0) = \frac{\gamma}{\Delta_C} \approx \frac{\vert \Psi(x_0)\vert^2 dV}{C}. 
\end{equation}
As in section \ref{mesj}, consider a large interval of time, say $T.$ The number of clicks at $x_0$ during this interval 
is given by
\begin{equation}
\label{EQj16C}
n_{T,C}(x_0) = \frac{T}{\Delta_C} \approx\frac{1}{\gamma C}  \vert \Psi(x_0)\vert^2 dV. 
\end{equation}
 Hence, the probability of detection at $x_0$ is 
\begin{equation}
\label{EQj17C}
P(x_0) = \frac{n_{T,C}(x_0)}{\int n_{T,C}(x) dx} \approx \frac{\vert \Psi(x_0)\vert ^2 dV}{\int \vert \Psi(x)\vert^2 dx}=
\vert \Psi(x_0)\vert ^2 dV.
\end{equation}
Surprisingly the probability of detection does not depend on calibration. 
(But the frequency of clicks decreases with increase of the scaling constant $C.$ Hence, if one increases
of the calibration level $C,$ then more time will be needed to collect enough clicks to obtain
a proper frequency estimation of the probability of detection.)

\section{No double clicks}
\label{CAL1}

We recall that Bohr elaborated his complementarity principle\footnote{This principle is often called
``wave-particle'' duality. However, we stress that Bohr had never used the latter terminology by himself.}
from analysis of the two slit-experiment. On the one hand, quantum systems exhibit interference properties
which are similar to properties of classical waves. On the other hand, these systems also exhibit particle 
properties. Wave properties (interference) are exhibited if both slits are open and experimenter does not 
try to control ``which slit passing''. At this experimental context one can be totally fine with a classical 
wave type model. However, if experimental context is changed and detectors are placed behind slits, then
``wave features of quantum systems disappear and particle features are exhibited.'' What does the latter mean?
Why is the usage of the wave picture impossible? Typically, it is claimed that, since 
classical wave is spatially extended, two detectors (behind both slits) can click simultaneoulsy and produce 
double clicks. However, as it is commonly claimed, there are no double clicks at all; hence, the wave 
model has to be rejected (in the context of the presence of detectors). Bohr had not find any reasonable
explanation of context dependent features of quantum systems and he elaborated the complementarity principle.

Of course, the cliam that there are no double clicks at all is meaningless at the experimental level. There 
are always double clicks. The question is whether the number of double clicks is very small (comparing 
with the numbers of single clicks). Corresponding  experiments have been done \cite{Grangier}, \cite{Grangier1} and it was
shown that the number of double clicks is relatively small. Such experiments are considered as confirmation
of Bohr's complementarity principle.

We show that the absence of double clicks might be not of the fundamental value, but a consequece of the procedure of calibration of detectors. Consider again a random signal $\phi.$ But now we take two threshold
type detectors located in neighbourhoods $V_{x_0}$ and $V_{y_0}$ of the points $x_0$ and $y_0.$ They are calibrated
with the connstant $C,$ see (\ref{TRESH}). For moments of clicks, we have two approximate equalities:
 \begin{equation}
\label{EQj11D}
\frac{1}{\gamma} \int_0^{\Delta_C(x_0)} \int_{V_{x_0}} \vert \phi(s, x)\vert^2dx ds \approx C\epsilon,
\end{equation}
 \begin{equation}
\label{EQj11D1}
\frac{1}{\gamma} \int_0^{\Delta_C(y_0)} \int_{V_{y_0}} \vert \phi(s, x)\vert^2  dx ds\approx C\epsilon,
\end{equation}
A double click corresponds to the (approximate) coincidence of moments of clicks
 \begin{equation}
\label{EQj11D2}
\Delta_C(x_0, y_0) = \Delta_C(x_0) =\Delta_C(y_0).
\end{equation}
Hence, by adding the approximate equalities (\ref{EQj11D}), (\ref{EQj11D1}) under condition (\ref{EQj11D2}) we obtain
 \begin{equation}
\label{EQj11DD}
\frac{1}{\gamma} \int_0^{\Delta_C(x_0, y_0)} \int_{V_{x_0}\cup V_{y_0}} \vert \phi(s, x)\vert^2dx ds \approx 2 C\epsilon,
\end{equation}
Again by using ergodicity and the assumption that the internal time scale of signals is essentially finer than
the time scale of measurement (``click production'') we obtain
$$
\frac{\Delta_C(x_0,y_0)}{\gamma}\Big[ \frac{1}{\Delta_C(x_0,y_0)} \int_0^{\Delta_C(x_0,y_0)} \int_{V_{x_0}\cup V_{y_0}} \vert \phi(s, x)\vert^2dx ds \Big] 
$$
$$
\approx \frac{\Delta_C(x_0,y_0)}{\gamma} 
\int_{V_{x_0}\cup V_{y_0}} \vert \psi(x)\vert^2 dx \approx 2 C\Vert \psi\Vert^2
$$
or, for normalized ``wave function'' $\Psi(x),$
$$
\frac{\Delta_C(x_0,y_0)}{\gamma} [ \int_{V_{x_0}} \vert \Psi(x)\vert^2 dx + 
\int_{V_{y_0}} \vert \Psi(x)\vert^2 dx]
$$
$$ =   \frac{\Delta_C(x_0,y_0)}{\gamma} [P(x\in V_{x_0}) +
P(x\in V_{y_0}] \approx 2 C.
$$ 
Hence
$$
P(\rm{double\; click}) = \frac{\gamma}{\Delta_C(x_0,y_0)} 
\approx \frac{1}{2C} [P(x \in V_{x_0}) +
P(x \in V_{y_0})]\leq \frac{1}{2C}.
$$
Hence, by increasing the calibration constant $C$ one is able to decrease the number of double clicks
to negligibly small. 

\section{Quantum probabilities from measurements of prequantum random fields}
\label{QP}

As was discussed in introduction, in a series of papers \cite{KH1}--\cite{KH1a} a purely wave model 
reproducing quantum averages was created -- PCSFT (prequantum classical statistical field theory). 
The main problem in matching PCSFT with
QM was violation of the spectral postulate of QM: prequantum physical varaibles are continuous quadratic forms
of prequantum signals, but in the realistic measurement procedures we have discrete clicks of detectors. 
We shall see that the aforementioned detection scheme can be coupled to PCSFT and the problem of discrete clicks 
can be solved.

\subsection{Essentials of prequantum classical statistical field theory}

Take Hilbert space $H$ as space of classical states.   Consider a probability
distribution $\mu$ on  $H$  having zero average (it means
that $$\int_H \langle y, \phi \rangle d\mu(\phi)=0$$ for any $y \in H.$) and
covariance operator $D$ which is  defined by a symmetric
positively defined bilinear form:
\begin{equation}
\label{CO} \langle D y_1, y_2 \rangle=\int_H \langle y_1, \phi \rangle \langle  \phi, y_2 \rangle d\mu(\phi),
y_1, y_2 \in H,
\end{equation}
By scaling we obtain operator
\begin{equation}
\label{COPI}
 \rho= D/\rm{Tr} D,
\end{equation}
 with $\rm{Tr} \rho=1.$
In PCSFT it is considered as a density operator.

By PCSFT a quantum state, a density operator, is simply the
symbolic representation  of the covariance operator  of the
corresponding prequantum (classical) probability distribution -- random field.

In general a probability distribution  is not
determined by its covariance operator. Thus the correspondence
PCSFT $\to$ QM is not one-to-one. However, if the class of
prequantum probability distributions  is restricted to Gaussian,
then this correspondence becomes one-to-one. 

In PCSFT classical variables are defined as functions from (Hilbert) state space $H$
to real numbers, $f=f(\phi).$ By PCSFT a quantum observable, a
self-adjoint operator, is simply a symbolic representation of $f$
by means of its {\it second derivative} (Hessian), $$f \to
\widehat{A}=\frac{1}{2} f^{\prime \prime}(0),$$ see \cite{KH1a} for detail. This correspondence
is neither one-to-one. However, by restricting the class of
classical variables to {\it quadratic forms} on Hilbert state space $H,$
$$f_A(\phi) = \langle \widehat{A} \phi, \phi \rangle,$$    we make  correspondence
PCSFT $\to$ QM one-to-one. 

And finally, we present the basic equality coupling the prequantum (classical random field) and quantum (operator) 
averages
\begin{equation}
\label{CO1} \langle f_A\rangle_\mu (\equiv E f_A(\phi)) =\int_M
f_A(\phi)  d\mu(\phi)= \rm{Tr} D \widehat{A}=
 (\rm{Tr} D) \;  \langle \widehat{A} \rangle_\rho.
\end{equation}
Thus the quantum average $\langle \widehat{A} \rangle_\rho$ can be
obtained as scaling of the classical average. We remark that
scaling parameter $\rm{Tr} D$ is, in fact, the dispersion of the
probability distribution $\mu:$
  \begin{equation}
\label{TR} \sigma^2\equiv E||\phi||^2 =
 \int_H ||\phi||^2 d\mu(\phi)={\rm Tr}\; D.
\end{equation}
It determines the scale of fluctuations of the prequantum random field.

\subsection{Born's rule from prequantum classical statistical field theory}

An important class of quantum states is given by pure states, normalized vectors $\Psi \in H.$ 
Any pure state $\Psi$ determines the density operator
$$
\rho= \vert \Psi\rangle \langle \Psi\vert.
$$

Take now a pure quantum state $\Psi,$ i.e., $\Vert \Psi \Vert=1.$ Consider a prequantum random field $\phi(x, \omega)$
(e.g., Gaussian) with zero average and the copvariance operator $D_\psi= \vert \psi\rangle\langle \psi\vert,$
where $$\psi= \sqrt{\epsilon} \Psi,$$ where $\epsilon>0$ is a small parameter such that $\epsilon^2$ has the physical dimension 
 of energy density. (The wave function of QM has the physical dimension of 1/volume.) Here
 $\epsilon= \int_H \Vert \phi \Vert^2 d \mu_\psi(\phi)$  is the average energy of the prequantum random field.
 In the process of measurement this field is considered as a random signal. 

Now we apply to this (so to say prequantum) random signal the scheme of 
 section \ref{mesj} and obtain the Born's rule. 

\medskip

Thus {\it we presented a model of discrete detection of 
 prequantum  random fields corresponding to quantum systems
 which reproduces the basic rule fo QM, the Born's rule.}

\section{The case of arbitrary covariance(density) operator}
\label{QP1}

We now generalize the above scheme of classical signal measurement to signals with arbitrary covariance operators.
(We recall that we had considered only  a very special class of signals with covariance operators 
given by one-dimensional projectors, see (\ref{eqg41}).) By applying this scheme to classical signals 
corresponding to quantum systems we obtain the Born's rule for quantum states given by density operators.
So, let $\phi(x, \omega)$ be a random signal with the covariance operator $D_\mu,$ where $\mu$ is the probability distribution
of this random signal, a normalized measure on the $L_2$-space. 
 Since this is a trace class operator
it can be represented as an integral operator:
\begin{equation}
\label{EQj16k1}
 D_\mu \phi(x) = \int_{{\bf R}^3} D_\mu(x,y) \phi(y) dy.
 \end{equation}
Let us consider the integral operator with the kernel 
\begin{equation}
\label{EQj16k2}
A_{x_0}(x,y) = \delta(x-x_0) \delta(y-x_0),
 \end{equation}
 where $x_0\in {\bf R}^3$ is a fixed point. Hence, 
 $\widehat{A}_{x_0} \phi(x)= \phi(x_0) \delta (x-x_0)$ and
\begin{equation}
\label{EQj16k3}
 \langle \widehat{A}_{x_0} \phi, \phi \rangle = 
\phi(x_0) \int \delta(x-x_0) \overline{\phi(x)} d x= 
\vert \phi(x_0)\vert^2. 
\end{equation}
 Set $\widehat{B}= D_\mu \widehat{A}_{x_0}.$ Then it has the kernel
 $$B(u,v)= \int D_\mu (u,z) A_{x_0} (z, v) dz= D_\mu (u,x_0) \delta(v-x_0).$$ Hence
 \begin{equation}
\label{EQj16k4}
\rm{Tr} \widehat{B}= \int B(u,u) du= D_\mu (x_0, x_0).
\end{equation}
Thus
 \begin{equation}
\label{EQj16k5}
 \int_H \vert \phi(x_0)\vert^2 d\mu(\phi) = \int_H \langle \widehat{A}_{x_0} \phi, \phi\rangle  d\mu(\phi)=
 \rm{Tr} D_\mu \widehat{A}_{x_0} = D_\mu (x_0, x_0).
\end{equation}
We also have 
 \begin{equation}
\label{EQj16k6}
\langle \pi\rangle =\rm{Tr} D_\mu = \int_{{\bf R}^3} D_\mu(x,x) dx.
\end{equation}

We now generalize the detection scheme of section \ref{mesj}. We can directly jump to the generalization of 
the relation (\ref{EQj13}):
\begin{equation}
\label{EQj16k7}
\frac{\Delta}{\gamma} D_\mu(x_0, x_0) dV \approx \int_{{\bf R}^3} D_\mu(x,x) dx,
\end{equation}
or
\begin{equation}
\label{EQj16k8}
\frac{\gamma}{\Delta} \approx \frac{D_\mu(x_0, x_0) dV}{\int_{{\bf R}^3} D_\mu(x,x) dx}.
\end{equation}
Thus the probability of signal detection at the point $x_0$ is given by 
\begin{equation}
\label{EQj16k0}
P(x_0) \approx \frac{D_\mu(x_0, x_0) dV}{\int_{{\bf R}^3} D_\mu(x,x) dx}.
\end{equation}

To find coupling with QM, we set $\rho= D_\mu/\rm{Tr} D_\mu.$ Then $\rm{Tr} \rho= 1$ and the above 
equality can be written in terms of the density operator $\rho:$
\begin{equation}
\label{EQj16k0}
P(x_0) \approx \rho(x_0, x_0) dV.
\end{equation}
This is nothing else than the Born's rule for the quantum state given by the density operator $\rho.$ 
Thus we demosntrated that even for a mixed quantum state it is possible to find such a prequantum random field
that the ``discrete-click'' measurement of this signal produces the basic probabilistic rule of QM.

\section{Discrete-counts  detection of classical random signals: general scheme}
\label{mesj1}

Let $e(x)$ be a fixed $L_2$-function.  We present a version of the discrete-detection scheme
of section \ref{mesj} by taking $e(x),$ instead of $\delta_{x_0} (x) = \delta(x-x_0).$ (In principle,
we may proceed without the restriction that $e \in L_2$ and select $e$ as a distribution; but we proceed 
with $L_2$-functions; on one hand this simplifies essentially mathematics of presentation; on the other hand,
in this section we are mainly interested in a signal-detection scheme underlying quantum measurements 
of observers with discrete spectra.) Consider a detector, ``$e$-antenna'', corresponding to the function 
$e(x).$ It works in the following way. As in the scheme of section \ref{mesj}, it integrates the energy of a signal during some interval of time, until the thershold $\epsilon$ is approached. Here $\epsilon$  is the average energy of 
a random signal. The only difference with section \ref{mesj} is that such a detector integrates the ``energy 
along the direction'' $e\in L_2,$ namely, $\frac{1}{\gamma} \int_0^\Delta \vert \langle \phi(s,x), e\rangle\vert ^2 ds.$
The rest of the detection scheme coincides with the one presented section \ref{mesj} (see also 
section \ref{QP1}):
\begin{equation}
\label{EQj16k11}
\frac{1}{\gamma} \int_0^\Delta \vert \langle \phi(s,x), e\rangle\vert^2 ds \approx \epsilon= \rm{Tr} D_\mu,
\end{equation}
where $D_\mu$ is the covarince operator of the probability distribution of a random signal under measurement.
\begin{equation}
\label{EQj16k12}
\frac{\Delta}{\gamma} \Big(\frac{1}{\Delta} \int_0^\Delta \vert \langle \phi(s,x), e\rangle\vert ^2 ds\Big) \approx \rm{Tr} D_\mu,
\end{equation}
\begin{equation}
\label{EQj16k13}
\frac{\Delta}{\gamma} \int_H \vert \langle \phi, e\rangle\vert^2 d\mu(\phi) \approx \rm{Tr} D_\mu,
\end{equation}
\begin{equation}
\label{EQj16k14}
\frac{\Delta}{\gamma} \langle D_\mu e, e \rangle \approx \rm{Tr} D_\mu,
\end{equation}
finally, for the probability of detection of this signal by the $e$-antenna, we get
\begin{equation}
\label{EQj16k15}
P(e) \approx \frac{\langle D_\mu e, e \rangle}{\rm{Tr} D_\mu}.
\end{equation}

Now to proceed to QM, we first normalize the covariance operator of the random signal, $\rho= D_\mu/\rm{Tr} D_\mu.$
Hence, 
\begin{equation}
\label{EQj16k16}
P(e) \approx\langle \rho e, e \rangle.
\end{equation}

Consider now a quantum observable $\widehat{A}$ with purely discrete nondegenerate spectrum; 
$\widehat{A}e_k = \lambda_k e_k.$   Consider the ensemble of detectors corresponding to the system of normalized eigenfunctions $\{e_k(x)\};$  ensemble of $e_k$-antennas. Each click induced by the $e_k$-antenna is associated with the value
$\lambda_k.$  Denote this classical observable  by the symbol $M_A.$
The average of $M_A$ is given by
\begin{equation}
\label{EQj16k17}
\langle M_A\rangle = \sum_k \lambda_k P(e_k)= \sum_k \lambda_k \langle \rho e_k, e_k \rangle.
\end{equation}
Thus we obtained the quantum formula for averages:
\begin{equation}
\label{EQj16k17}
\langle M_A\rangle = \rm{Tr} \rho \widehat{A}. 
\end{equation}

We now can extend our approach (representation of quantum measurements as 
``click-measurements'' of classical random signals) to quantum observables 
with discrete degenerate spectra. Consider a subspace $L$ of $H$ and the corresponding 
$L$-antenna. Let $\widehat{P}_L$ be the corresponding orthogonal projector. The $L$-antenna
produces a click if 
\begin{equation}
\label{EQj16k18}
\frac{1}{\gamma} \int_0^\Delta \Vert \widehat{P}_L \phi(s)\Vert^2 ds \approx \epsilon= \rm{Tr} D_\mu,
\end{equation}
\begin{equation}
\label{EQj16k18}
\frac{\gamma}{\Delta}  \approx \rm{Tr} D_\mu \widehat{P}_L /\rm{Tr} D_\mu,
\end{equation}
i.e., 
\begin{equation}
\label{EQj16k18}
P(L) \approx \rm{Tr} D_\mu \widehat{P}_L /\rm{Tr} D_\mu,
\end{equation}
where $\rho$ is the normalization of the covariance operator of the signal. 

Now take a quantum observable $\widehat{A}$ with discrete in general degenerate spectrum;
consider its eigenspaces $L_k$  corresponding  eigenvalues $\lambda_k$ Then measurements of 
this quantum observable can be modeled through measurements performed by a system of 
$L_k$-antennas.

In this approach there is no reason to consider orthogonal projectors and self-adjoint operators; a POVM,
$\{\widehat{Q}_k\}, \sum_k \widehat{Q}_k=I, \widehat{Q}_k \geq 0,$ 
arises naturally as the symbolic representation of  generalization of aforementioned scheme (of measurement of 
classical random signals) to the family of $Q_k$-antennas: each integrates the $Q_k$-transform of a classical signal.
The starting point of generalization is the realtion:
\begin{equation}
\label{EQj16k18}
\frac{1}{\gamma} \int_0^\Delta \Vert \widehat{Q}_k \phi(s)\Vert^2 ds \approx \epsilon= \rm{Tr} D_\mu.
\end{equation}

By starting with Dirac's approach to quantum observables (based on generalized eigenvectors) 
we can easily generalize our scheme to operators with continuous spectra and, in particular, obtain
the position measurement as a special case. 

This paper was written under support of the grant QBIC (2008-2011), Tokyo University of Science, and the grant
Mathematical Modeling, Linnaeus Unievrity (2008-2010).

\end{document}